\documentclass[prb,superscriptaddress,reprint,amsmath,amssymb,aps]{revtex4-2}
\usepackage{graphicx}
\usepackage{dcolumn}
\usepackage{bm}
\usepackage{xcolor}
\usepackage{booktabs}
\usepackage{hyperref}

\newcommand{\moire}{moir\'{e}}

\begin{document}

\title{Effects of Lithium Intercalation in Twisted Bilayer Graphene}

\author{Daniel T. Larson}
\affiliation{Department of Physics, Harvard University, Cambridge, Massachusetts, 02138, USA.}

\author{Stephen Carr}
\affiliation{Department of Physics, Harvard University, Cambridge, Massachusetts, 02138, USA.}

\author{Georgios A. Tritsaris}
\affiliation{John A. Paulson School of Engineering and Applied Sciences, Harvard University, Cambridge, Massachusetts, 02138, USA.}

\author{Efthimios Kaxiras}
\affiliation{Department of Physics, Harvard University, Cambridge, Massachusetts, 02138, USA.}
\affiliation{John A. Paulson School of Engineering and Applied Sciences, Harvard University, Cambridge, Massachusetts, 02138, USA.}

\date{\today}

\begin{abstract}
We investigate the effects of lithium intercalation in twisted bilayers of 
graphene, using first-principles electronic structure calculations. To model this system we employ commensurate supercells that correspond to twist angles of 7.34$^\circ$
and 2.45$^\circ$. From the energetics of lithium absorption we demonstrate that for low Li concentration the intercalants cluster in the AA regions with double the density of a uniform distribution. The charge donated by the Li atoms to the graphene layers results in modifications to the band structure that can be qualitatively captured using a continuum model with modified interlayer couplings in a region of parameter space that has yet to be explored either experimentally or theoretically. Thus, the combination of intercalation and twisted layers simultaneously provides the means for spatial control over material properties and an additional knob with which to tune \moire{} physics in twisted bilayers of graphene, with potential applications ranging from energy storage and conversion to quantum information.
\end{abstract}

\maketitle

\section{Introduction}
Interest in the physics of two-dimensional (2D) layered materials has grown significantly since the successful isolation of graphene layers in 2004~\cite{novoselov2004electric}. More recent work has started to explore the emergent properties of assemblies in which a few such layers are stacked on top of each other in a controllable manner, being held together by weak van der Waals forces~\cite{geim2013van}. An additional degree of freedom in stacked 2D layers is the relative rotation between adjacent layers, which leads to formation of \moire{} super-lattices. The variation of the local
atomic environment associated with the \moire{} patterns has the potential to create interesting electronic
effects on scales larger than that of the atomic lattice~\cite{bistritzer_moire_2011,cao_unconventional_2018}.
The study of the variation of the electronic structure as a function of the relative rotation, or ``twist", is essential in exploring the implications for applications in fields ranging from energy storage and conversion to quantum information.

Small atoms and molecules have often been used to 
probe charge transfer and the associated energy 
landscape in nanostructures \cite{bediako_heterointerface_2018,larson2018lithium,tritsaris_diffusion_2012, lu_using_2012}. 
Intercalated compounds of graphene constitute 
derivative systems in which novel physics may emerge. Lithium atoms are
known to preferentially intercalate in AA-stacked graphene
bilayers~\cite{shirodkar2016li}, so it is natural to expect that in a twisted
bilayer graphene system the Li intercalants will preferentially
congregate in the highly localized AA regions. This leads to local
charge doping of the graphene layers, with interesting effects on the electronic band structure. Insight into the origin of magic angles and flat bands for empty bilayers has been gained by studying a chirally symmetric continuum model~\cite{tarnopolsky2019origin}, where the interlayer coupling in the AA regions is taken to zero. In the case of intercalated bilayers, the interlayer coupling near the AA regions is \emph{increased} compared to the AB/BA regions, which tunes the system into the opposite parameter regime that remains to be explored, both theoretically and experimentally.

In this paper we use first-principles calculations based on Density Functional Theory (DFT) to demonstrate the clustering of intercalated Li atoms in the AA regions of a graphene bilayer with a 2.45$^\circ$ twist. Then, by comparing the DFT band structure to a continuum model with adjustable interlayer coupling terms, we gain insight into the nontrivial effects that the Li intercalants have on the electronic states in the \moire{} pattern.  

The paper is organized as follows: Section~\ref{sec:methods} describes the twisted graphene unit cells and DFT parameters. Section~\ref{sec:EI} discusses the energetics of Li absorption in twisted bilayers of graphene and describes our prediction for Li clustering at low concentrations. Section~\ref{sec:structure} examines in detail structural changes in the bilayer that accompany Li absorption. Section~\ref{sec:electronic} compares the DFT band structures modified by Li intercalation with modified continuum model band structures for empty bilayers, and Section~\ref{sec:conclusion} presents our conclusions and points to the next stages of exploration.

\section{Computational Methods}
\label{sec:methods}
We begin with optimized lattice parameters for a primitive graphene
unit cell and then build two ideal, commensurate, twisted bilayer
supercells. The smaller cell, with a 7.34$^\circ$ twist,
contains 244 C atoms; the larger cell, with a 2.45$^\circ$ twist,
contains a total of 2188 C atoms. To describe the various distances in each supercell we adopt the
notation of~\cite{zhang2018structural} and let $L$ be the size of the
\moire{} unit cell. For the small angles we are considering,
$L\approx \sqrt{3} a/\theta_0$ where $a=1.42$ \AA{} is the
carbon-carbon bond length and $\theta_0$ is the initial twist
angle. We further define $\ell = L/\sqrt{3}$, which is the distance
between the center of an AA region and a neighboring AB (or BA)
region, and take $R_{AA}$ to be the radius of the AA region (as defined
and computed in Ref.~\cite{zhang2018structural}). These various lengths are
collected in Table~\ref{tab:lengths} for several angles $\theta_0$.

For total energy calculations we use DFT based on the Projector Augmented Wave (PAW) formalism, as implemented in the
VASP~\cite{Kresse1996,Kresse1996a} code. We elected to use the PBE
exchange-correlation functional~\cite{Perdew1996}, and a plane-wave
energy cutoff of 400 eV. Van der Waals corrections were included using
the DFT-D3 method~\cite{Grimme2010}. Using a single $k$-point, the ionic positions were relaxed until the magnitude of all forces was less than 0.01
eV/\AA. The ground state energies and band structures were
computed from the relaxed positions using a $\Gamma$-centered
3$\times$3 $k$-point grid which has 5 $k$-points in the
irreducible Brillouin zone, including the high symmetry point $K$.

\begin{table}
  \centering
\begin{tabular}{|c|c|c|c|c|}\hline
$\theta_0$ (degrees) & $L$ (\AA) & $R_{AA}$ (\AA) & $R_{AA}/\ell$ & \# C atoms
  \\ \hline
  7.34 & 19.2 & $\sim$5 & $\sim$0.5     & 244 \\
  2.88 & 47.0 & 12.7 & 0.47 & 1588 \\
  2.45 & 56.4 & 14.6 & 0.45 & 2188 \\
  2.01 & 70.5 & 17.0 & 0.42 & 3268 \\
  1.47 & 93.9 & 19.6 & 0.36 & 6076 \\
  0.99 & 140.9 & 21.7 & 0.27 & 13,468\\
  0.50 & 281.8 & 22.5 & 0.14 & 53,068\\
  \hline
\end{tabular}
\caption{Dependence of the \moire{} cell size, $L$ (\AA), radius of the AA
  region, $R_{AA}$ (\AA), ratio $R_{AA}/\ell$, and number of C atoms on the initial twist
  angle $\theta_0$ (degrees). The values of $R_{AA}$ were taken from
  Figure 4(b) of Ref.~\cite{zhang2018structural}.}
\label{tab:lengths}
\end{table}

\section{Energetics of Lithium Intercalation}
Li atoms intercalating between twisted layers of graphene balance the preference for the low-energy carbon environment of AA regions against the repulsive force of other Li atoms. In this section we demonstrate that for low Li concentrations they will cluster in the AA regions with roughly double the density of a uniform distribution, that is, a distribution of the \emph{same} number of Li atoms uniformly spread throughout the entire unit cell.

\label{sec:EI}
\subsection{Single Lithium Intercalant}

In order to compare the energy profile for Li intercalants in \moire{} cells corresponding to different twist angles, we measure the location
of the Li atom in units of $\ell$, the distance between the AA and AB regions. For the 7.34$^\circ$ cell $\ell = 11.1$ \AA. To survey the energy landscape, a single Li ion was placed in 11 different initial locations between the layers and each structure was allowed to fully relax.

For the larger supercell with a 2.45$^\circ$ twist angle, $\ell =
33.2$ \AA. In this case the energy landscape was mapped out by
relaxing the structure with a single Li atom located at each of 14 locations along the line connecting an AA region with an AB region, and also at each of 7 locations along one half of the line connecting two AA
regions. The latter direction is the domain wall (DW) that separates an AB
region from a BA region. In each case the entire structure 
was allowed to fully relax.
\begin{figure*}
\centering
\includegraphics[width=\textwidth]{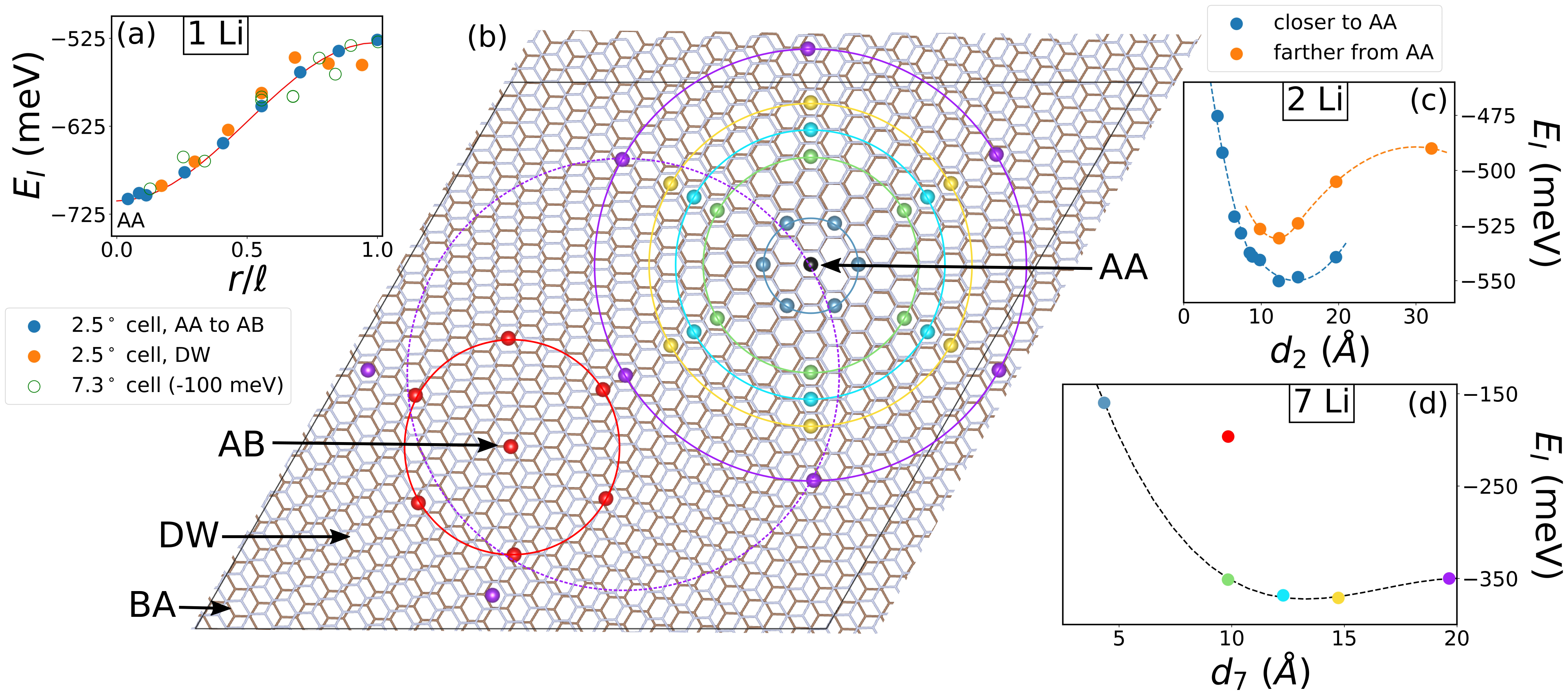}
\caption{(a) Intercalation energy, $E_I$, for single Li intercalants in the 2.45$^\circ$ cell (filled circles) and 7.34$^\circ$ cell (open green circles). Blue points represent intercalants along the line connecting AA and AB regions; orange points are along the domain wall connecting AA regions. The green points have been shifted by $-100$ meV for ease of comparison. (b) Configurations of 6 Li atoms arranged in a circle around a 7th Li atom (black) located at the center of an AA region in the 2.45$^\circ$ graphene cell. The Li-Li distances, $d_7$, are: 4.3 (blue), 9.8 (green), 12.3 (cyan), 14.7 (yellow), and 19.7 (purple) \AA. For $d_7 = 19.7$ \AA{} the Li atoms from neighboring AA regions are almost as close as the neighbors centered around the same AA region, as indicated by the dashed purple circle. The red circle represents a configuration with $d_7=9.8$ \AA, centered at an AB region. (c) Average intercalation energy per atom, $E_I$, for pairs of Li intercalants, as a function of the distance between the Li atoms, $d_2$. The orange circles represent configurations where one Li atom is very close to the center of the AA region but the other atom is much farther away, whereas the blue points represent configurations where the two Li atoms are more evenly spaced about the center of the AA region. The lines are simple fits as guides to the eye. (d) $E_I$ vs $d_7$ for the configurations of 7 Li atoms shown in (b). The colors of the data points match the colors of the atoms and circles in (b). The line is a simple polynomial fit.}
\label{fig:combined}
\end{figure*}

We define the average intercalation (binding) energy per lithium atom as follows:
\begin{equation}
E_{I}=\frac{1}{n} \left[ E({\rm Li}_n{\rm Gr})- E({\rm Gr}) - n E({\rm
    Li})\right]. 
\end{equation}
Here $E(\rm{Gr})$ is the total energy of the empty graphene bilayer
and $E(\rm{Li}) = -2.001$ eV is the energy of a single Li
atom in bulk bcc Li, calculated using similar VASP parameters. Smaller (more negative) values indicate stronger binding.

The intercalation energy for a single Li atom is plotted in
Fig.~\ref{fig:combined}(a) as a function of $r/\ell$, where $r$ is the
distance from the center of the AA region. The filled circles give intercalation energy of a Li atom in the 2.45$^\circ$ cell located along the line connecting an AA region to an AB region (blue) and along the domain wall connecting two AA regions (orange). The open green circles show $E_I$ for the 7.34$^\circ$ cell; they have been shifted by $-100$ meV to emphasize the similarity of the $r/\ell$ dependence. All three cases show a similar monotonic increase in intercalation energy when moving away from the AA region, which can be fit by a simple cosine function that describes the data remarkably well: $f(r/\ell) = [-620 - 90 \cos (\pi r/\ell)]$ meV. These specific calculations suggest that there is a
universal energy profile that depends only on the fractional distance
from the AA region and is independent of the twist angle, with an
overall energy scale of 180 meV. This substantiates a simple physical
picture in which the Li intercalant is primarily sensitive to the
local arrangement of carbon atoms, with energy increasing as the local
environment shifts away from perfect AA stacking.

This universal behavior cannot be expected to hold once the
twist angle becomes too small. The absolute size of the AA region
increases as the twist angle is decreased to roughly
1$^\circ$, after which point the size of the AA region remains
relatively constant with decreasing twist angle due to relaxation effects~\cite{zhang2018structural}. On the other hand, the overall \moire{}
length scale continues to increase as $\sim 1/\theta$, and thus the
\emph{relative} size of the AA region decreases significantly. Using $R_{AA}$
extracted from Fig.~4(b) of Ref.~\cite{zhang2018structural}, we can
calculate the ratio $R_{AA}/\ell$, shown in
Table~\ref{tab:lengths}. For angles between 3.0$^\circ$ and
2.0$^\circ$ this ratio does not change much, decreasing from 47\% to
42\%, which suggests that the universal energy profile inferred from
the 7.34$^\circ$ and 2.45$^\circ$ calculations will likely hold down to
twist angles of 2.0$^\circ$. For a twist angle of 1.5$^\circ$, $R_{AA}/\ell$
has decreased to 36\%, so it is unlikely that the shape of the energy
profile will be the same as for larger angles. We expect the lateral size of the higher energy portion of the profile to grow at the expense of the lower energy portion, while maintaining the overall scale of the energy difference between AA and AB regions.

\subsection{Li-Li interactions}

Lithium intercalants are sensitive not only to the environment of the graphene bilayer host, but also to the presence of other nearby Li atoms. Fig.~\ref{fig:combined}(c) shows the variation in the intercalation energy for pairs of Li intercalants as a function of the distance $d_2$ between them. In this plot the orange points represent configurations where one Li atom is very close to the center of the AA region but the other atom is much further away, whereas the blue points represent configurations where the two Li atoms are more evenly spaced about the center of the AA region. The repulsion between the positively charged Li intercalants, which have donated electrons to the nearby carbon atoms (as discussed in Section~\ref{sec:electronic}), is readily apparent in the decrease of $E_I$ with increasing separation out to
$d_2 \sim 12$ \AA. For larger $d_2$, the
effect of the local graphene environment becomes important and $E_I$
begins to increase again as the second Li atom is necessarily farther
from the preferred region of AA stacking. For two Li atoms, these
competing effects can be balanced by splitting the distance from AA
between the two Li atoms, as demonstrated by the lower energy of the blue points compared to the orange points for $d_2\geq 10$ \AA{}. This balance will become more complicated
as additional Li ions enter the bilayer.

In order to study the potential for clustering of Li intercalants in
the AA regions we consider the intercalation energy for
configurations of 7 Li atoms, where one is at the center of
an AA region and the other 6 are arranged in a circle of radius $d_7$
around it such that
the distance between all nearest Li neighbors is the same. We have
studied 5 different values of $d_7$, as shown in
Fig.~\ref{fig:combined}(b), and also one configuration with $d_7=9.8$
\AA{} centered at an AB region (red). 
The region occupied by each configuration is indicated by a colored circle that matches the colors of the atoms. The dashed purple circle of radius $d_7=19.7$ \AA{} in Fig.~\ref{fig:combined}(b) shows that at this separation the Li distribution is nearly uniform, since atoms
centered at one AA region are roughly the same distance from Li atoms centered at other AA regions (purple atoms in the lower left of the \moire{} cell). The intercalation energy per Li atom, $E_I$, for each of these configurations is shown in Fig.~\ref{fig:combined}(d) as a function of $d_7$, with data points colored to match the atoms and circles in Fig.~\ref{fig:combined}(b). The minimum for $d_7=14.7$ demonstrates that the
potential well around the AA region arising from the \moire{} pattern
is sufficient to produce clustering in those regions for
concentrations of lithium at least up to 7 Li intercalants per
\moire{} cell. For that value of $d_7$ the local Li concentration at the AA location is
Li$_1$C$_{144}$, or $n_\mathrm{Li} = 5.3\times10^{13}$ cm$^{-2}$,
which is over twice what the concentration would be if the same number of Li ions
were uniformly distributed over the entire cell. $E_I$ for the
configuration centered on the AB site (red) is over 150 meV higher than $E_I$ of the same
configuration centered on the AA site, which is
another demonstration of lithium's preference for the AA regions.

It is interesting to compare this concentration to experimental
results for Li intercalation into untwisted bilayer graphene. Based on
their measurements, the authors of Ref.~\cite{ji2019lithium} infer a
lithium intercalation process with identifiable steps at
concentrations of Li$_1$C$_{244}$, Li$_1$C$_{85}$, Li$_1$C$_{28}$, and
Li$_1$C$_{12}$, the latter corresponding to full intercalation. Our
estimated concentration in the AA region of the twisted bilayer,
namely Li$_1$C$_{144}$, is intermediate between the first two steps of
the proposed process, which is to be expected due to the confining
potential of the AA region which will act to concentrate the Li atoms
more than in a uniform bilayer.

As the twist angle decreases below 2.45$^\circ$, the increasing size of
the AA regions will permit more Li atoms to cluster within them at the
same concentration. It is
interesting to note that for a 2.45$^\circ$ twist, the lowest energy cluster of
7 Li atoms has a preferred radius of $d_7 = 14.7$ \AA, which is very
close to the radius of the AA region, $R_{AA}=14.6$ \AA, as shown in
Table~\ref{tab:lengths}. Since the depth of the confining potential
around the AA regions can be reasonably assumed to be independent of
the twist angle (being the difference in energy between an AA and AB
intercalation site), one would expect the \emph{concentration} of Li atoms
that can be confined in the AA region to also be independent of twist
angle. Thus we can estimate the number of Li atoms that can cluster in
an AA region based on the preferred distance between them in the
2.45$^\circ$ cell, $d_7=14.7$ \AA. For angles below $\sim 1^\circ$ the
size of the AA region is constant, with a radius of $\sim 22$
\AA. A triangular lattice of Li atoms with lattice constant $\sim
14.7$ \AA{} could accommodate $\sim19$ Li atoms within each AA region.

\section{Structural Changes}
\label{sec:structure}
\begin{figure}[h]
\centering
\includegraphics[width=\columnwidth]{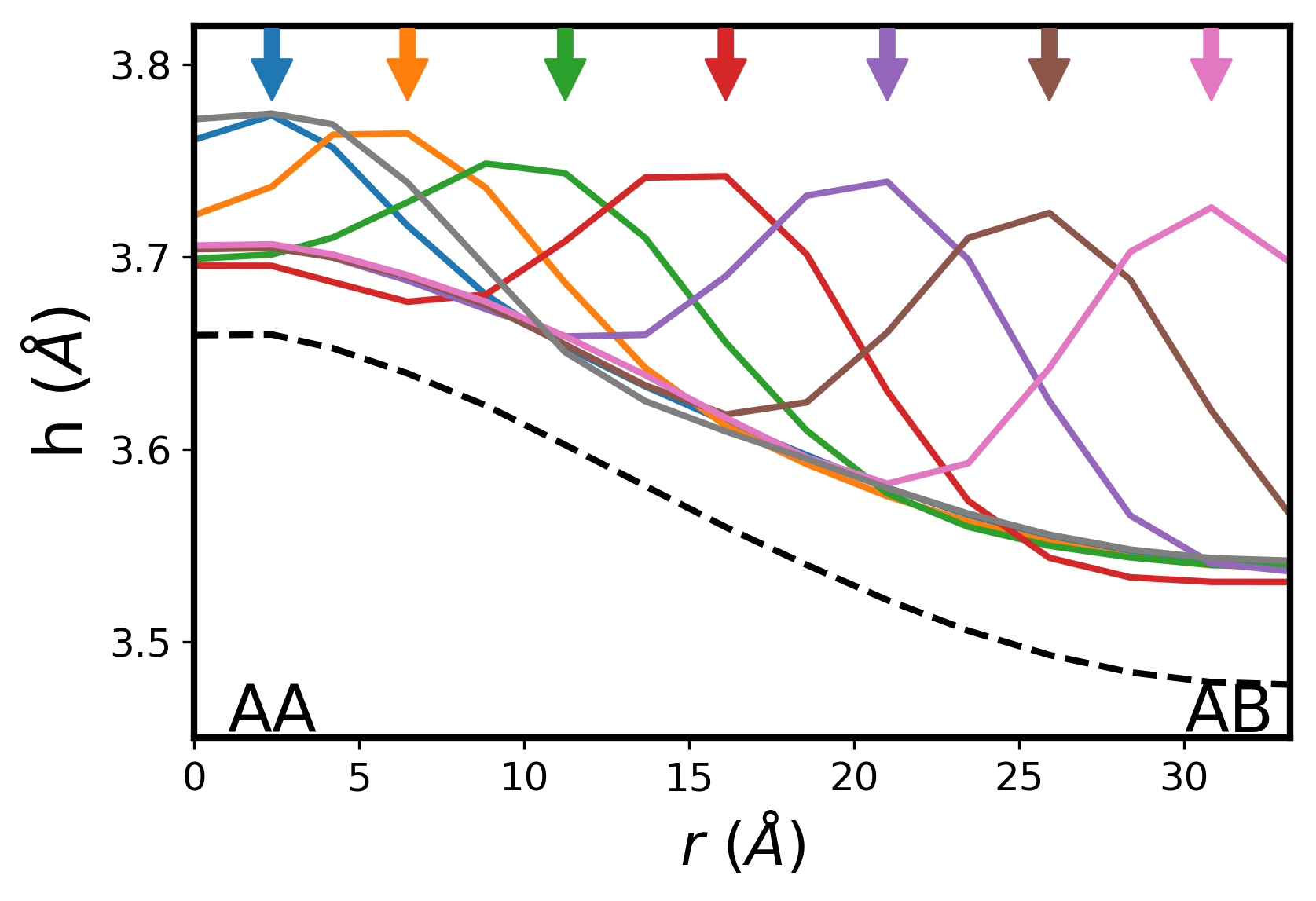}
\caption{Average separation $h$ (\AA) between graphene layers in a 5 \AA{}
  circle centered at points a distance $r$ (\AA{}) along a line between AA and AB regions. Colored curves show the separation in a cell with a Li
  intercalant located a fixed distance from the AA region, marked with a matching colored arrow above. The grey curve shows the separation when 7 Li atoms are concentrated
  within 5 \AA{} of the AA region. The dashed black curve 
shows the layer separation at the same locations in the empty bilayer.}
\label{fig:layer_sep}
\end{figure}

Intercalation of Li results in structural changes to the bilayer. Insertion of a single Li atom increases the
average layer spacing $h$ from 3.54 to 3.59 \AA, for all 14 Li
locations along the line from AA to AB. Fig.~\ref{fig:layer_sep}
shows how the overall separation increases compared to the empty
bilayer (dashed black line), and also how the local separation is
significantly increased near the location of a Li atom. The effect of
the Li on layer separation is largest near AB, where the layers are
closest together in the empty bilayer. Additional Li intercalants
appear to broaden the spatial extent of the increased layer
separation, but do not further increase the distance between the
layers, as shown by the grey curve in
Fig.~\ref{fig:layer_sep}, which gives the local separation for a
cell with 7 Li atoms all clustered within 5 \AA{} of the AA center.

As discussed in Ref.~\cite{zhang2018structural}, the local rotation angle $\theta_l$ of the AA region can differ from the initial rotation angle $\theta_0$ due to relaxation effects. The presence of Li increases the interlayer distance, consequently decreasing the interlayer interaction and presumably reducing the impact of relaxation on the local structure. We can explore this effect by calculating the local rotation angle $\theta_l$ from our DFT results: $\theta_l = \Delta x/r_{AA}$, where $\Delta x$ is the in-plane distance between a chosen pair of carbon atoms that would be vertically aligned in the untwisted structure, and $r_{AA}=7.1$ \AA{} is the in-plane distance of that chosen pair from the center of rotation at the middle of the AA region. The results are shown in Table~\ref{tab:angle}. The unrelaxed structure has a local rotation angle $\theta_l = \theta_0 = 2.45^\circ$, whereas the fully relaxed, empty bilayer has $\theta_l=2.67$, which is similar to the angles reported in Ref.~\cite{zhang2018structural}, namely $\theta_l=2.8^\circ$  corresponding to $\theta_0=2.4^\circ$. The addition of Li does seem to reduce the local rotation angle closer to $\theta_0$, indicating a reduction in the relaxation by as much as 50\% for the case with 7 Li atoms.

\begin{table}
  \centering
\begin{tabular}{|c|c|c|}\hline
  \#Li & comment & $\theta_l$ (deg)\\ \hline
    0 & relaxed & 2.67 \\ 
      1 & AA & 2.61 \\
  1 & AB & 2.61 \\
  7 & $d_7=19.7$ & 2.57 \\
  0 & unrelaxed & 2.45 \\
\hline
\end{tabular}
\caption{Final local rotation angle $\theta_l$ in the AA region for a cell
  with initial rotation angle $\theta_0 = 2.45^\circ$, calculated
  from a pair of carbon atoms $r_{AA}=7.1$ \AA{} from the center of
  rotation. 
  }
\label{tab:angle}
\end{table}

\section{Effects on Electronic Structure}
\label{sec:electronic}

\begin{figure}
  \includegraphics[width=\columnwidth]{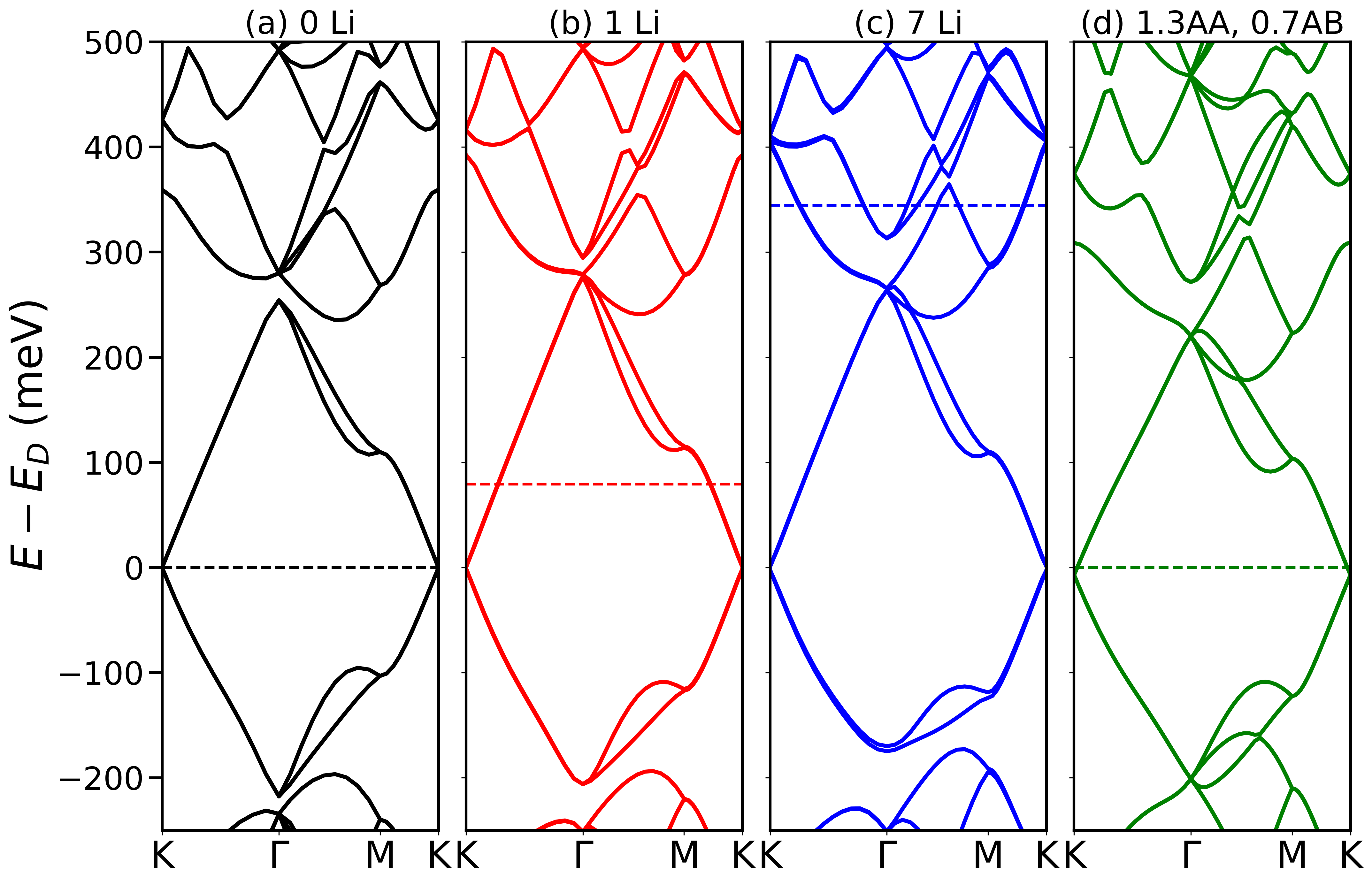}
\caption{(a)-(c) DFT band structures of the 2.45$^\circ$ twisted bilayer graphene cell with 0, 1, and 7 Li intercalants in the AA region. The locations of the Li atoms have been chosen to minimize the total energy, with Li-Li distances approximately $15$ \AA{}. (d) Continuum model with 30\% increase (decrease) in the AA (AB) interlayer coupling. In each case the Fermi level is indicated by a dashed line of the same color.}
\label{fig:2.5bands}
\end{figure}
Having determined the distribution of Li ions under conditions of low intercalation, we can then study their effects on the electronic structure. Fig.~\ref{fig:2.5bands} shows the band structure for the 2.45$^\circ$  twisted bilayer with 0, 1, and 7 Li intercalants in the AA region. Calculations with 3 and 5 Li intercalants show that the evolution of the bands is a smooth, monotonic function of the number of Li atoms. The locations of the Li atoms have been chosen to minimize the total energy, which results in Li-Li distances of approximately 15 \AA{}. The band structure of the empty bilayer calculated directly with DFT agrees with that calculated using a tight-binding-based continuum model~\cite{carr2019exact,fang2019angle}. Already with a single Li intercalant at the AA site we see a qualitative change in the symmetry of the conduction bands at $\Gamma$. While the empty structure has a doubly degenerate energy level below a four-fold-degenerate level at $\Gamma$, that pattern becomes inverted upon Li intercalation, with the four-fold degenerate level appearing at lower energies than the two-fold degenerate level. This pattern is reminiscent of the band topology that appears for twist angles well below the magic angle, as shown in Fig.~2(d) of Ref.~\cite{yoo2019atomic}. The intercalation of Li makes this set of electronic states accessible at much larger twist angles. Moreover, Li donates charge carriers that allow the Fermi level to be tuned through these lowest states at $\Gamma$. 

By using the continuum model with modified interlayer coupling terms we have determined that this band inversion occurs once the AA interlayer coupling becomes roughly 20\% larger than the AB interlayer coupling. In Fig.~\ref{fig:2.5bands}(d) we plot the band structure of an empty 2.45$^\circ$ cell calculated using the continuum model where the AA / AB interlayer couplings have been adjusted by $\pm$30\% respectively. Focusing first on the energy window between 200 and 300 meV above the Dirac point, we see that an increase in the AA interlayer coupling relative to the AB interlayer coupling results in bands that are in good qualitative agreement with the DFT results for 7 Li atoms clustered near the AA region. On the other hand, for bands below the Dirac point, the Li intercalated band structure does not show any inversion of the band order, though it does exhibit an increased gap between groups of bands at $\Gamma$ in the -200 to -100 meV range. Adjusting the interlayer coupling in the opposite sense ($\pm$30\% for AB / AA, respectively) does not cause band inversion in the 200 to 300 meV range, but the gaps at $\Gamma$ increase in both the 200-300 meV range and in the -200 to -100 meV range. In Ref.~\cite{tarnopolsky2019origin} the authors study the chirally symmetric continuum model, the limit of vanishing AA interlayer coupling, to gain insight into the appearance of magic angles and flat bands. They further study the magic angles and band gap as a function of the ratio of AA to AB interlayer couplings in the range of 0 to 1, where the experimental value of this ratio is approximately 0.7-0.8. The theoretical approach is to fix a ratio of interlayer couplings and then find the angle where the Fermi velocity vanishes or the band width is minimized. On the other hand, in experiments the twist angle is the control parameter and the ratio of interlayer couplings changes with the angle (and related structural relaxation), presumably with the AA interlayer coupling decreasing compared to AB coupling as the twist angle approaches the first magic angle from above. In this work we have shown that Li intercalants offer a new way to adjust the ratio of interlayer couplings by increasing the AA coupling compared to the AB coupling, at least for states near the Fermi energy. This motivates theoretical study of the continuum model for the ratio of couplings greater than 1. But because the continuum model is nearly particle-hole symmetric, no single set of parameters can accurately approximate the very non-trivial effect that Li intercalants have on the band structure, which motivates experimental work to explore these new parameter ranges in the twisted bilayer parameter space.

\begin{figure}
  \includegraphics[width=\columnwidth]{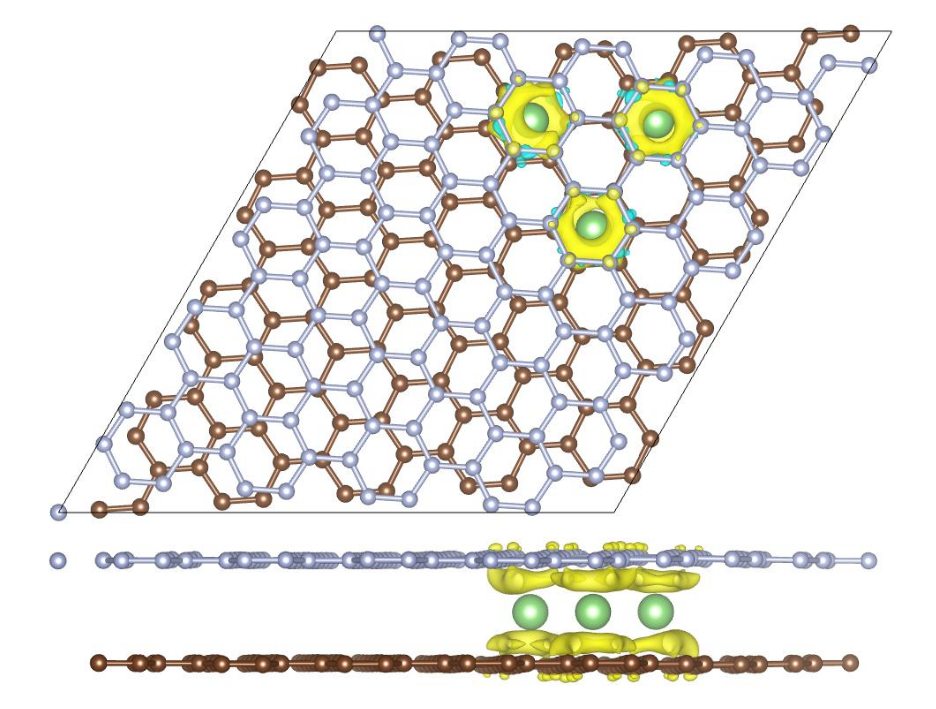}
\caption{Top and side views of the charge density difference for the
  7.34$^\circ$ cell with 3 Li atoms located near the AA region. Yellow
clouds represent excess electrons.}
\label{fig:chg-diff}
\end{figure}

To verify that the changes to the band structure arise from the localized charge donated by the Li intercalants we calculated the Bader charge distribution for both the 7.34$^\circ$
and 2.45$^\circ$ twisted bilayer cells, with various numbers of Li
intercalants. In all cases the Li atoms lose $\sim$0.85 electrons
each, with the excess charge localized in the nearest 12 carbon atoms, that is, a hexagon of 6 C in each layer. As a specific example,
Fig.~\ref{fig:chg-diff} shows the charge density difference
$\Delta\rho = \rho(\mathrm{Gr+3Li}) - \rho(\mathrm{Gr}) - \rho(\mathrm{3Li})$ for the 7.34$^\circ$ cell with 3 Li atoms near the AA region. Excess electrons are
shown in yellow. According to the Bader partitioning, the Li atoms lose
a total charge of 2.6$e$, with 0.2$e$ assigned to the vacuum. The 36
atoms closest to the Li atoms (6 C surrounding each Li in each
layer) gain a total of 2.7$e$, while the rest of the C atoms together
lose 0.3$e$.
Furthermore, DFT calculations for the empty 2.45$^\circ$ twisted cell with 7 additional electrons and compensating uniform background charge show no changes in the bands, demonstrating the key role that the charge localization has in modifying the energy levels.

\section{Conclusions}\label{sec:conclusion}
We used first-principles calculations based on 
DFT to explore the
energetic, structural, and electronic changes associated with the intercalation of Li between two graphene layers in
commensurate systems with small twist angles of 7.34$^\circ$ and 2.45$^\circ$. We found that the local potential wells at the AA regions
in the \moire{} pattern of twisted bilayer graphene are sufficiently
deep ($\sim$180 meV) to overcome the Li-Li repulsion and induce
clustering in the AA regions, at least for low Li concentrations. For
the $2.45^\circ$ twist angle studied, Li intercalation at an
average concentration of $n\sim 2.45\times10^{13}$ cm$^{-2}$ would
result in clustering in the AA regions, producing a local Li density
approximately double the average value. This clustering of intercalants and the resulting localized charge transfer to the graphene layers results in significant qualitative changes to the electronic band structure. Using a continuum model with modified interlayer couplings, we were able to show that the changes in the band degeneracies at the $\Gamma$-point can result from an increase of the AA interlayer coupling compared to the AB interlayer coupling. In contrast, the states below the Dirac point behave as if the AA interlayer coupling were decreased relative to the AB coupling, highlighting the complicated interplay between intercalants and \moire{} physics.

\section*{Acknowledgments}
We thank Dr. Tamar Mentzel and Mehdi Rezaee for useful discussions.
Partial support by the DOE BES Award No. DE-SC0019300 is acknowledged.
This work used the Stampede2 supercomputer at the Texas Advanced
Computing Center through allocation TG-DMR120073, which is part of the
Extreme Science and Engineering Discovery Environment (XSEDE),
supported by NSF Grant No. ACI-1548562. We also used the Odyssey
cluster supported by the FAS Division of Science, Research Computing
Group at Harvard University, and resources of the Argonne Leadership
Computing Facility, which is a DOE Office of Science User Facility
supported under Contract DE-AC02-06CH11357.

\bibliography{twLi}

\end{document}